\documentclass[letter, scriptaddress, twocolumn, prd, showpacs,showkeys]{revtex4-1}

	\usepackage{amsmath}
        \usepackage{mathrsfs}
	\usepackage{amssymb}
	\usepackage{graphicx} 
	\usepackage{makeidx}
	\usepackage{amsfonts}
	\usepackage[ansinew]{inputenc}
	\usepackage[usenames, dvipsnames]{pstricks}
	\usepackage{epsfig}
	\usepackage{pst-grad} 
	\usepackage{pst-plot} 
	\usepackage[colorlinks, hyperindex]{hyperref}
	\hypersetup
	{
		colorlinks, %
		citecolor=blue, %
		linkcolor=Green, %
		urlcolor=blue, %
	}

\makeindex

\begin{document}

\title{\large{{\bf Unitary Black hole radiation: Schwarzschild-global monopole background}}}
\author{Arpit Das}
\email{ad13ms118@iiserkol.ac.in}
\author{Narayan Banerjee}
\email{narayan@iiserkol.ac.in}
\affiliation{Department of Physical Sciences, \\
Indian Institute of Science Education and Research Kolkata, \\
Mohanpur, West Bengal 741246, India.}

\begin{abstract}
Black hole radiation from an infinitesimally thin massive collapsing shell, possessing a global monopole charge, which in turn leads to a Schwarzschild black hole with a global monopole charge has been shown to be processed by a unitary evolution. The exterior metric of the collapsing shell is described by the global monopole (GM) metric. The analysis is performed using the Wheeler-deWitt formalism which gave rise to a Schr\"{o}dinger-like wave equation. Existence of unitarity is confirmed from two independent lines of approach. Firstly, by showing that the trace of the square of the density matrix, of the outgoing radiation, from a quantized massless scalar field, is unity. Secondly, by proving that the conservation of probability holds for the wave function of the system.  
\end{abstract}


\maketitle
\section{Introduction}
Recently, in an attempt to shed some light on the resolution of the information loss paradox \cite{hawking1, hawking2, wallace1, mathur1, polchinski1, page1}, it has been shown by Das and Banerjee\cite{das1} that radiation from a collapsing charged shell is processed with a unitary evolution. This was achieved in a Reissner-Nordstr\"om background using the Wheeler-deWitt formalism\cite{dewitt, wheeler} and unitarity checks were carried out using two independent lines of approach, density matrix and conservation of probability. We extend the result as given in \cite{das1} by performing the same kind of analysis for a not asymptotically flat spacetime. We adopt the formalism and method of analysis from \cite{das1} and apply it to a global monopole background metric \cite{barriola1}. It was shown in \cite{dadhich1} that a Schwarzschild black hole with a global monopole charge Hawking radiation is Planckian in nature. So, naturally it is a relevant theoretical question to investigate unitarity issues in such backgrounds. This is the primary motivation of this work. \\

The present work shows that the process of black hole radiation, in a not asymptotically flat spacetime, is unitary. Saini and Stojkovic\cite{saini2} worked with a not asymptotically flat spacetime before, specifically with an asymptotically AdS spacetime. However, the results obtained therein are based on numerical estimates. For not asymptotically flat spacetimes, our analysis and therby the results obtained from them are more robust as they are done analytically. \\

We work with a metric that includes a global monopole charge $\eta$. The Schwarzschild case as considered in \cite{saini1}, is recovered trivially as a special case by setting $\eta=0$. \\

In section 2 we describe the global monopole metric. Section 3 contains the description of the model. The scalar field is discussed in section 4. The unitarity is ascertained in section 5. The last section includes a discussion of the results.

\section{The Global Monopole}
The metric for a Schwarzschild black hole with a global monopole charge $\eta$ is given in natural units as, \cite{barriola1, dadhich1, konstantin1},
\begin{align}
ds^2_{GM} = &-\left(1-\eta^2-\frac{2M}{r}\right)dt^2 + \left(1-\eta^2-\frac{2M}{r}\right)^{-1}dr^2 \nonumber\\
&+ r^2d\Omega^2_2, \label{metric1}   
\end{align}
where, $\eta^2<<1$ and $M$ is the mass of the black hole. Note that the above metric is not asymptotically flat and even with $M=0$ the spacetime is not flat, as it has some non-zero curvature \cite{dadhich1},
\begin{align}
&R^0_0 = R^1_1 = 0  = R_{01}, \label{ricci1} \\    
and, \ &R^2_2 \propto \frac{\eta^2}{r^2}, \label{ricci2}
\end{align}
where the above terms are the components of the Ricci tensor. The observational signature of a global monopole is in the existence of a ``solid angle deficit''.\\
\\
The event horizon is at,
\begin{align}
R_{GM} = \frac{2M}{1-\eta^2}.    
\end{align}
$\left.\right.$\\
Let us also give below the stress-energy tensor corresponding to the Global monopole field \cite{konstantin1},
\begin{align}
T^0_0 = T^1_1 = \frac{\eta^2}{8\pi r^2} ,   
\end{align}
where we see that the total energy is divergent and so solutions of such form as $eq^n(\ref{metric1})$ are unrealistic and perhaps appear in some instances of cosmic phase transition \cite{barriola1}.
\\
\\
The surface gravity $\kappa$ for the metric as given in $eq^n(\ref{metric1})$ is obtained by noting that the metric is of the form \cite{poisson1},
\begin{align}
&ds^2 = -fdt^2 + f^{-1} dr^2 + r^2d\Omega^2_2, \label{f1} \\
implying, \ &\kappa = \frac{f^{'}(r)}{2}, \label{k1} \\
implying, \ &\kappa_{GM} = \frac{\left(1-\eta^2\right)^2}{4M} \label{k2} \\ 
&\left(as, \ f(r) = 1-\eta^2-\frac{2M}{r}\right). \nonumber
\end{align}
where $\kappa_{GM}$ is the surface gravity for the global monopole metric. \\
\\
The semi-classical study of the metric as given in $eq^n(\ref{metric1})$ was done in \cite{dadhich1} and it was show that the outgoing Hawking radiation is thermal possessing a Planck spectrum,
\begin{align}
N = \frac{1}{e^{8\pi M\omega/(1-\eta^2)^2}-1}    
\end{align}
where $N$ is the number density of outgoing quanta of particles. The Hawking temperature is recovered to be,
\begin{align}
T_{GM} = \frac{\left(1-\eta^2\right)^2}{8\pi M}, \label{T1}    
\end{align}
which can also be obtained from $eq^n(\ref{k2})$ using the Hawking relation $T_{H}=\frac{\kappa}{2\pi}$ (which holds here too).

\section{The Model}
In our model we have an infinitesimally thin massive collapsing spherical shell with a global monopole charge \cite{dadhich1, konstantin1}, whose background metric is $g_{\mu\nu}$. There is also a massless scalar field $\Phi$ whose dynamics we shall study. We assume that $\Phi$ couples to the gravitational field (which originates from the presence of a non-trivial background metric). However, $\Phi$ does not directly couple to the shell. An asymptotic observer, at the future null infinity, is present to detect the outgoing flux with a detector and by assumption does not interact with the ``shell-metric-scalar'' system. Hence, the observer does not significantly affect the evolution of the system and similarly for the system vis-a-vis the observer. The action for the whole system is then given by \cite{stojkovic1},
\begin{align}
S_{tot} = &\int d^4x\sqrt{-g}\left[-\frac{\mathcal{R}}{16\pi}+\frac{1}{2}(\partial_\mu\Phi)^2\right]-\sigma\int d^3\xi\sqrt{-\gamma} \nonumber\\ 
&+ S_{obs}, \label{Stot}
\end{align} 
where the first term denotes the usual Einstein-Hilbert term for the background metric $g_{\mu\nu}$, the second term represents the action for the massless scalar field, the third term represents the shell's action in terms of its world-volume coordinates $\xi^a(a=0,1,2)$, $\sigma$ is the tension of the shell (or, the shell's proper energy density per unit surface area) and $\gamma_{ab}$ is the shell's induced world-volume metric, given by,
\begin{align}
\gamma_{ab} &= g_{\mu\nu}\partial_a X^\mu\partial_b X^\nu, \label{gamma}
\end{align}
where $X^\mu(\xi^a)$ determines the location of the shell. The Roman indices run over the internal world-volume coordinates $\xi^a(a=0,1,2)$ while the Greek indices run over the usual spacetime coordinates.\\

The last term $S_{obs}$ represents the action for the observer. 

\subsection{Spacetime Foliation-GM coordinates}
The mass and the global monopole charge is confined in an infinitesimally thin shell \cite{konstantin1}, as per our considerations. So that for an exterior observer the distribution would be spherical. However, the inside of the shell would be empty and would be described by the Minkowski metric. The exterior of the shell is described by a Global Monopole metric. Thus, we have,
\begin{align}
ds^2_{out} = &-\left(1-\eta^2-\frac{2M}{r}\right)dt^2 \nonumber \\ 
&+ \left(1-\eta^2-\frac{2M}{r}\right)^{-1}dr^2 + r^2d\Omega_2^2,   \label{metric-out}\\
ds^2_{in} = &-dT^2 + dr^2 + r^2d\Omega_2^2,  \label{metric-in}\\
ds^2_{on-shell} = &-d\tau^2 + r^2d\Omega_2^2,  \label{met-onshell}
\end{align}
for $r>R(t)$, $r<R(t)$ and $r=R(t)$ respectively. Here $r$ is the radial coordinate. So $r=R(t)$ describes the collapsing shell and $R:=R(t)$ is the radius of the shell. $T$, $\tau$ and $t$ are the time coordinate inside the shell, proper time on the shell and time coordinate of the exterior observer respectively. $d\Omega^2_2$ is the standard $S^2$ metric. \\
\\
An important consideration to observe here is that since the above GM coordinates would lead to a coordinate singularity, at $R=R_{GM}$ (the event horizon), we might face trouble using this for our analysis. However, observe that from the point of view of an asymptotic observer, the event horizon is an infinitely red shifted surface. So, the observer can only observe the collapse of the shell approaching its event horizon in infinite time as per his time $t$. Thus, the analysis would happen upto this limit which is relevant from an asymptotic viewpoint and the GM coordinates are well behaved upto this limit, that is just outside the event horizon.
\\
\\
Similar to \cite{das1}, we consider timelike unit vectors $u^\alpha := \frac{dx_{out}^\alpha}{d\tau}$ and $v^\alpha := \frac{dx_{in}^\alpha}{d\tau}$, for $d s^2_{out}$ and $ds^2_{in}$ respectively. From their normalization, that is, $u^\alpha u_\alpha=-1$ and $v^\alpha v_\alpha=-1$, one obtains, at $r=R(t)$, $t_{\tau} = \frac{\sqrt{E+R_{\tau}^2}}{E}$, $T_{\tau} = \sqrt{1+R_{\tau}^2}$ and $T_t = \sqrt{E-(1-E)\frac{R_t^2}{E}}$. In the above expressions, a subscript indicates a differentiation w.r.t. that particular coordinate. $x^\alpha_{out}$ and $x^\alpha_{in}$ are the coordinates pertaining to $ds^2_{out}$ and $ds^2_{in}$ respectively. Also, $E:=1-\eta^2-\frac{2M}{R(t)}$.

\subsection{Mass of the shell}
According to Israel's formulation\cite{israel1, israel2, konstantin1}, the mass $M$ of the shell can be obtained as,
\begin{align}
M &= 4\pi\sigma R^2\left[\sqrt{1+R_\tau^2}-2\pi\sigma R\right]-\frac{\eta^2R}{2}, \label{M}
\end{align}
We shall show below that $M$ would turn out to be a constant of motion. So, there would be no conflict with the fact that $M$ is a constant of integration in the metric and can be identified as the mass of the shell. Similar to the results given in \cite{lopez1}, one can write,
\begin{align}
&\frac{R_{\tau\tau}}{\alpha} = \frac{\eta^2}{8\pi\sigma R^2}+6\pi\sigma-\frac{2\alpha}{R}, \ \left(where, \ \alpha:=\sqrt{1+R_\tau^2}\right). \label{M_c}
\end{align}
Now, using $eq^ns$(\ref{M}) and (\ref{M_c}), 
\begin{align*}
M_\tau = & \ R_\tau\left[8\pi\sigma R(\alpha-2\pi\sigma R)-\frac{\eta^2}{2}\right] \\
&+R_\tau\left[4\pi\sigma R^2\left(\frac{\eta^2}{8\pi\sigma R^2}-\frac{2\alpha}{R}+6\pi\sigma-2\pi\sigma\right)\right] =0.
\end{align*}
Thus, we see that $M$ is a constant of motion.\\

Since, we have proven that $M$ is a constant of motion, we can have the following identification,
\begin{align}
\mathcal{H}_{shell}\equiv M, \label{H-s}
\end{align}
where $\mathcal{H}_{shell}$ is the Hamiltonian of the shell. $\mathcal{H}_{shell}$ is to be treated classically for our analysis.

\subsection{Action for the shell}
The shell's action is given as,
\begin{align}
S_{shell} = -\int dT \ \left[4\pi\sigma R^2\left[\sqrt{1-R_T^2}-2\pi\sigma R\right] - \frac{\eta^2R}{2}\right]. \label{S-shell}
\end{align}

The Lagrangian corresponding to the shell's action yields the conjugate momentum as,
\begin{align}
\Pi_{shell} = \frac{\partial\mathcal{L}_{shell}}{\partial R_T} = 4\pi\sigma R^2\left(\frac{R_T}{\sqrt{1-R_T^2}}\right). \label{P-shell}
\end{align}

Now the Hamiltonian is,
\begin{align}
\mathcal{H}_{shell} = & \ \Pi_{shell}R_T - \mathcal{L}_{shell} \nonumber\\
= & \ 4\pi\sigma R^2\left[\sqrt{1+R_\tau^2}-2\pi\sigma R\right]-\frac{\eta^2R}{2}.
\label{H-shell}
\end{align}
$\mathcal{H}_{shell}$ as obtained above matches with $M$ as expressed in $eq^n(\ref{M})$. Hence, the action in $eq^n$(\ref{S-shell}) is consistent (since, this action gives the correct $\mathcal{H}_{shell}$ as expressed in $eq^n$(\ref{H-s})). Now let us consider $S_{shell}$ in terms of time $t$, (using the expression for $T_t$),
\begin{align}
S_{shell} = &-\int dt \ \left[4\pi\sigma R^2\left[\sqrt{E-\frac{R_t^2}{E}}\right]\right] \nonumber\\
&+\int dt \ \left[4\pi\sigma R^2\left[2\pi\sigma R\sqrt{E-\frac{1-E}{E}R_t^2}\right]\right] \nonumber\\
&+\int dt \ \left[\frac{\eta^2R}{2}\sqrt{E-\frac{1-E}{E}R_t^2}\right]. \label{S-st}
\end{align}
Let us also consider the conjugate momentum and Hamiltonian in terms of $t$,
\begin{align}
\Pi_{shell} = & \ \frac{\partial\mathcal{L}_{shell}}{\partial R_t} \nonumber\\ 
= & \ \frac{4\pi\sigma R^2R_t}{\sqrt{E}}\left[\frac{1}{\sqrt{E^2-R_t^2}}-\frac{2\pi\sigma R(1-E)}{\sqrt{E^2-(1-E)R_t^2}}\right] \nonumber\\
&- \frac{4\pi\sigma R^2R_t}{\sqrt{E}}\left[\frac{\eta^2(1-E)}{8\pi\sigma R\sqrt{E^2-(1-E)R_t^2}}\right], \label{P-st}
\end{align}

\begin{align}
\mathcal{H}_{shell} = & \ \Pi_{shell}R_t - \mathcal{L}_{shell} \nonumber\\ 
= & \ 4\pi\sigma E^{3/2}R^2\left[\frac{1}{\sqrt{E^2-R_t^2}}-\frac{2\pi\sigma R}{\sqrt{E^2-(1-E)R_t^2}}\right] \nonumber\\
&- 4\pi\sigma E^{3/2}R^2\left[\frac{\eta^2}{8\pi\sigma R\sqrt{E^2-(1-E)R_t^2}}\right]. \label{H-st}
\end{align}

\subsection{Incipient Limit}
We define the so-called incipient limit, $R\rightarrow R_{GM}$, as the limit when the radius of the shell approaches the event horizon. From $eq^n$(\ref{P-st}) and $eq^n$(\ref{H-st}) we note that, as $R\rightarrow R_{GM}$, 
\begin{align}
&\Pi_{shell} = \frac{4\pi\mu R^2 R_t}{\sqrt{E}\sqrt{E^2-R_t^2}}, \label{P-sti}
\end{align}
\begin{align}
&\mathcal{H}_{shell} = \frac{4\pi E^{3/2}\mu R^2}{\sqrt{E^2-R_t^2}}, \label{H-sti}
\end{align}
where, $\mu:=\sigma\left(1-2\pi\sigma R_{GM}-\frac{\eta^2}{8\pi\sigma R_{GM}}\right)$. Then we have,
\begin{align}
\mathcal{H}_{shell} = [(E\Pi_{shell})^2+E(4\pi\mu R^2)^2]^{1/2} \equiv [q^2+m^2]^{1/2}, \label{H-strel}
\end{align}
where $q^2:=(E\Pi_{shell})^2$ and $m^2:=E(4\pi\mu R^2)^2$.\\
\\
$\mathcal{H}_{shell}$ as given in $Eq^n$(\ref{H-strel}), is the Hamiltonian of a relativistic particle with a position dependent mass. This is how the shell behaves in the incipient limit as $R\rightarrow R_{GM}$. We shall show below that in this limit also, $\mathcal{H}_{shell}$ would turn out to be a constant of motion. Since, $\frac{d\mathcal{H}_{shell}}{d\tau}=\frac{\partial \mathcal{H}_{shell}}{\partial\tau}$, we have,
\begin{align}
&\frac{d}{d\tau}\left(4\pi\mu\frac{E^{3/2}R^2}{\sqrt{E^2-R_t^2}}\right) = 0 \nonumber\\
leading \ to, \ & \frac{E^{3/2}R^2}{\sqrt{E^2-R_t^2}} = \frac{\mathcal{H}_{shell}}{4\pi\mu} =: h \ (a \ constant), \label{h} \\
&(as \ \tau \ doesn't \ appear \ explicitly \ in \ \mathcal{H}_{shell}). \nonumber
\end{align}

These expressions can be arrived at independently using an alternative approach (see appendix).\\

Classically, we have from $eq^n$(\ref{h}) and from the expression of $T_t$,
\begin{align}
&R_t = \pm E\sqrt{1-\frac{E R^4}{h^2}} \approx \pm E\left(1-\frac{1}{2}\frac{ER^4}{h^2}\right) \approx \pm E \label{R-ti} \\
&(as \ R\rightarrow R_{GM}), \nonumber\\
&T_t = E\sqrt{1+(1-E)\frac{R^4}{h^2}}, \label{T-ti}
\end{align}
where solving $eq^n(\ref{R-ti})$ in terms of $t$ will give us the classical behaviour of the shell as $R(t)\rightarrow R_{GM}$. \\

$E$ can be written as,
\begin{align}
E = (1-\eta^2)\left(1-\frac{R_{GM}}{R}\right) = \epsilon\left(1-\frac{R_{GM}}{R}\right), \label{D-i1}
\end{align}
where $\epsilon:=(1-\eta^2)$.\\
\\
In the incipient limit, $E\rightarrow 0$ (as $R(t)\rightarrow R_{GM}$). Then, in this limit, $R_t\approx\pm E$. Now solving for $R(t)$ we get (from $eq^n$(\ref{R-ti}) and $eq^n$(\ref{D-i1})),
\begin{align}
\pm 1=\frac{1}{\epsilon}\frac{R}{R-R_{GM}}\frac{dR}{dt} &\approx \frac{1}{\epsilon}\frac{R_{GM}}{R-R_{GM}}\frac{dR}{dt} \nonumber\\
&(upto \ leading \ order) \nonumber\\
integrating, \ & R_{GM} ln\left(\frac{R_f-R_{GM}}{R_0-R_{GM}}\right) = \pm \epsilon t_f \nonumber\\ 
&(R_0:=R(0) \ and \ R_f:=R(t_f)) \nonumber\\
thus, \ & R_f = R_{GM} + (R_0-R_{GM}) \ e^{\pm \epsilon t_f/R_{GM}}, \label{Rf-ti1}
\end{align}
where the lower limit of integration w.r.t. $t$ is $t=0$ and the upper limit is $t=t_f$.\\
\\
Similar to as in \cite{das1}, as $R_f\rightarrow R_{GM}$ and $t_f>0$ along with $\epsilon>0$ (as, $\eta^2<<1$), we observe that, $t_f\rightarrow\infty$. Thus, the negative sign for $R(t)$ describes a collapsing model in the incipient limit. $Eq^n$(\ref{Rf-ti1}) also shows that from the viewpoint of an asymptotic observer, the formation of the event horizon takes infinite time implying that the event horizon is an infinite red shifted surface, which matches with the classical result, as stated earlier while choosing the GM coordinates. 

\section{The scalar field $\Phi$}
The action for the scalar field $\Phi$ can be written as a sum of the actions,
\begin{align}
S_{\Phi} =& \left.S_{\Phi}\right)_{in} + \left.S_{\Phi}\right)_{out} \nonumber \\
= & \ 2\pi\int dt\left[-(\partial_t\Phi)^2\left(\int_0^R dr \ r^2\frac{1}{T_t}\right)\right] \nonumber\\
&+2\pi\int dt\left[(\partial_r\Phi)^2\left(\int_0^R dr \ r^2 \ T_t\right)\right] \nonumber\\
&+2\pi\int dt\left[-(\partial_t\Phi)^2\left(\int_R^\infty dr \ r^2\frac{1}{1-\eta^2-\frac{2M}{r}}\right)\right] \nonumber\\
&+2\pi\int dt\left[(\partial_r\Phi)^2\left(\int_R^\infty dr \ r^2\left(1-\eta^2-\frac{2M}{r}\right)\right)\right], \label{S-phicom}
\end{align}
where the limits of the integration w.r.t. $r$ for $S_{\Phi})_{in}$ are from $0$ to $R$ and for $S_{\Phi})_{out}$ are from $R$ to $\infty$.
\\
\\
$T_t\rightarrow E \ (upto \ leading \ order)$ in the incipient limit (from $eq^n$(\ref{T-ti})). Thus,
\begin{align*}
\lim_{R\rightarrow R_{GM}}\frac{T_t}{1-\eta^2-\frac{2M}{r}}=\frac{R-\eta^2 R-2M}{r-\eta^2 r - 2M}\frac{r}{R} = 0.
\end{align*}
$T_t$ vanishes faster than $\left(1-\eta^2-\frac{2M}{r}\right)$ in the limit $R\rightarrow R_{GM}$. Thus, for the coefficients of $-(\partial_t\Phi)^2$, the $\frac{1}{T_t}$ term dominates. For the coefficients of $(\partial_r\Phi)^2$, the dominating term is $\left(1-\eta^2-\frac{2M}{r}\right)$. Therefore, in the incipient limit,
\begin{align}
S_{\Phi} \rightarrow & \ 2\pi\int dt\left[-\frac{1}{E}\int_0^{R_{GM}}dr \ r^2 (\partial_t\Phi)^2\right] \nonumber\\
&+2\pi\int dt\left[\int_{R_{GM}}^{\infty}dr \ r^2 \left(1-\eta^2-\frac{2M}{r}\right)(\partial_r\Phi)^2\right]. \label{S-phii}
\end{align}

\subsection{Mode expansion for $\Phi$}
For $\Phi$, one can easily check from its equation of motion, that is $\partial^2\Phi = 0$, that for $r<R(t)$ (from $\left.S_{\Phi}\right)_{in}$),
\begin{align}
\frac{\partial^2\Phi}{\partial r^2} + \frac{2}{r}\frac{\partial\Phi}{\partial r} = \frac{1}{T_t^2}\frac{\partial^2\Phi}{\partial t^2}-\frac{T_{tt}}{T_t^3}\frac{\partial\Phi}{\partial t}, \label{wavein}
\end{align}
where $T_t$, along with its powers and derivatives w.r.t. $t$, are independent of $r$.\\
\\ 
Similarly, for $r>R(t)$, we have (from $\left.S_{\Phi}\right)_{out}$)),
\begin{align}
&\left(1-\eta^2-\frac{2M}{r}\right)^2\frac{\partial^2\Phi}{\partial r^2} \nonumber\\
&+ \frac{2(r-M)}{r^2}\left(1-\eta^2-\frac{2M}{r}\right)\frac{\partial\Phi}{\partial r} = \frac{\partial^2\Phi}{\partial t^2}. \label{waveout}
\end{align}
From $eq^n$(\ref{wavein}) and $eq^n$(\ref{waveout}), we notice the following mode expansion (due to the separability property satisfied by the above equations),
\begin{align}
\Phi(r,t) = \sum_k a_k(t)f_k(r), \label{phi-rt}
\end{align}
where $a_k(t)$ are the modes and $f_k(r)$ are some real-valued smooth functions of r.
\\
\\
Now $S_\Phi$ in terms of modes $a_k$ is (as $R\rightarrow R_{GM}$),
\begin{align}
S_\Phi = \int dt \ \sum_{k,k^{'}} \left[-\frac{1}{2E}\frac{da_k}{dt}A_{kk^{'}}\frac{da_{k^{'}}}{dt}+\frac{1}{2}a_kB_{kk^{'}}a_{k^{'}}\right], \label{S-phim}
\end{align}
where $A_{kk^{'}}$ and $B_{kk^{'}}$ are defined as,
\begin{align}
&A_{kk^{'}} := 4\pi\int_0^{R_{GM}}dr \ r^2f_k(r)f_{k^{'}}(r), \label{Akk} \\
&B_{kk^{'}} := 4\pi\int_{R_{GM}}^{\infty}dr \ r^2\left(1-\eta^2-\frac{2M}{r}\right)f^{'}_k(r)f^{'}_{k^{'}}(r), \label{Bkk}
\end{align}
where, $f_k^{'}(r) := \frac{\partial f_k(r)}{\partial r}$. Observe that, both $A_{kk^{'}}$ and $B_{kk^{'}}$ are independent of $r$ and $t$ (as no $R(t)$ appears in them). \\
\\
Following \cite{das1}, we define the conjugate momenta, $\pi_k$s (to the modes $a_k$) as,
\begin{align}
\pi_k := \frac{\partial \mathcal{L}_{\Phi}}{\partial \dot{a_k}} \equiv -i\frac{\partial}{\partial a_k}, \label{pk}  
\end{align}
where $\dot{a}_k := \frac{da_k}{dt}$, and from $eq^n$(\ref{S-phim}), we have (with $\mathcal{L}_{\Phi}$ defined as the Langrangian for $\Phi$),
\begin{align}
\mathcal{L}_{\Phi} &= \sum_{k,k^{'}} \left[-\frac{1}{2E}\dot{a}_k A_{kk^{'}}\dot{a}_{k^{'}}{dt}+\frac{1}{2}a_kB_{kk^{'}}a_{k^{'}}\right], \label{L-phim1} \\
\mathcal{L}_{\Phi} &= -\frac{1}{2E}(\mathbf{\dot{a}}^T \mathbf{A}\mathbf{\dot{a}}) + \frac{1}{2}(\mathbf{a}^T \mathbf{B}\mathbf{a}), \label{L-phim2} 
\end{align}
where $\mathbf{A}$ and $\mathbf{B}$ are non-singular linear operators, such that, $A_{kk^{'}}\in\mathbf{A}$ and $B_{kk^{'}}\in\mathbf{B}$ in the chosen bases, say $\lbrace \dot{a_k}\rbrace$ and $\lbrace a_k\rbrace$ respectively. In the basis $\lbrace a_k\rbrace$, $\mathbf{a}$ is a column vector, such that, $a_k\in\mathbf{a}$. One can similarly express $\mathbf{\dot{a}}$ in the basis $\lbrace \dot{a_k}\rbrace$.\\
\\  
For the Hamiltonian of $\Phi$, $\mathcal{H}_{\Phi}$, we obtain,
\begin{align}
\mathcal{H}_{\Phi} = &\sum_{k} \pi_k\dot{a}_k - \mathcal{L}_{\Phi}\nonumber\\ 
= &\sum_{k,k^{'}} \left[\frac{1}{2E}\dot{a}_k A_{kk^{'}}\dot{a}_{k^{'}}{dt}+\frac{1}{2}a_kB_{kk^{'}}a_{k^{'}}\right] \label{H-phim1} \\
= & \ \frac{E}{2}(\mathbf{\Pi}^T\mathbf{A}^{-1}\mathbf{\Pi})+\frac{1}{2}(\mathbf{a}^T \mathbf{B}\mathbf{a}), \label{H-phim2}
\end{align}
where $\mathbf{\Pi}$ is a column vector, such that, $\pi_k\in\mathbf{\Pi}$, in a chosen basis say $\lbrace \pi_k\rbrace$ and $\mathbf{A}^{-1}$ is the inverse of $\mathbf{A}$.\\
\\ 
Following arguments similar to \cite{das1}, note that, $\mathbf{B}$ and $\mathbf{A}$ are real and symmetric infinite dimensional matrices and hence are self-adjoint. Therefore, by the {\itshape{Spectral Theorem}}, there exists orthonormal bases of position space and momentum space consisting of respective eigenvectors of $\mathbf{B}$ and $\mathbf{A}$. Furthermore, all the corresponding eigenvalues are real. Say, for instance, the bases for position space and momentum space are $\lbrace b_k\rbrace$ and $\lbrace \dot{b}_k\rbrace$ respectively (where, each $b_k$ is a linear combination of the original basis vectors $a_k$ and each $\dot{b}_k$ is a linear combination of the original basis vectors $\dot{a}_k$).

\subsection{The Schr\"{o}dinger-like wave equation}
If we study the equation for one eigenvector $b\in\lbrace b_k\rbrace$, then our conclusion will be the same for all other eigenvectors (see \cite{stojkovic1}). So, we shall solve the Schr\"{o}dinger-like wave equation for a wave functional $\Psi(\lbrace b_k\rbrace,t)$, which by the above assumption of equivalence is now a wave function $\psi(b,t)$. Therefore, $\psi(b,t)\equiv\Psi(\lbrace b_k\rbrace,t)$. Hence, using $eq^n$(\ref{L-phim2}), we write the Schr\"{o}dinger-like wave equation (for a single eigenvector $b$) as,
\begin{align}
\left[-\left(1-\eta^2-\frac{2M}{R}\right)\frac{1}{2\alpha}\frac{\partial^2}{\partial b^2}+\frac{1}{2}\beta b^2\right]\psi(b,t) = i\frac{\partial \psi(b,t)}{\partial t}, \label{Schr-t}
\end{align}
where, $\alpha$ and $\beta$ are the eigenvalues of $\mathbf{A}$ and $\mathbf{B}$ respectively.\\ 
\\
We define a new time parameter,
\begin{align}
&\widetilde{\eta} := \int_0^t dt \ \left(1-\eta^2-\frac{2M}{R}\right) \label{eta1} \\
leading \ to, \ &\frac{\partial \widetilde{\eta}}{\partial t} = E, \label{eta2}
\end{align}
and write $eq^n$(\ref{Schr-t}) as
\begin{align}
\left[-\frac{1}{2\alpha}\frac{\partial^2}{\partial b^2}+\frac{\beta}{2E}b^2\right]\psi(b,\widetilde{\eta}) = i\frac{\partial \psi(b,\widetilde{\eta})}{\partial \widetilde{\eta}}. \label{Schr-eta}
\end{align}
$Eq^n$(\ref{Schr-eta}) becomes,
\begin{align}
\left[-\frac{1}{2\alpha}\frac{\partial^2}{\partial b^2}+\frac{1}{2}\alpha\omega^2(\widetilde{\eta})b^2\right]\psi(b,\widetilde{\eta}) = i\frac{\partial \psi(b,\widetilde{\eta})}{\partial \widetilde{\eta}}, \label{Schr-sho}
\end{align}
where, we have chosen to set $\widetilde{\eta}(t=0)=0$ and $\omega$ is defined as,
\begin{align}
\omega^2(\widetilde{\eta}) := \left(\frac{\beta}{\alpha}\right)\frac{1}{E} =:\frac{\omega_0^2}{E}. \label{omega} 
\end{align}

We observe that, $eq^n(\ref{Schr-sho})$ is a time dependent Simple Harmonic Oscillator (SHO) equation with $\omega(\widetilde{\eta})$ as the frequency. \\

In the incipient limit (using  $eq^n(\ref{D-i1})$ and  $eq^n(\ref{R-ti})$),  
\begin{align}
\frac{dE}{dt} &= \frac{2M}{R^2}\frac{dR}{dt} = \epsilon\frac{2M}{\epsilon R^2}\frac{dR}{dt} \approx -\epsilon E\frac{R_{GM}}{R_{GM}^2} = -\frac{\epsilon E}{R_{GM}}. \ \ \  \label{Dt-i}
\end{align}
Integrating $eq^n(\ref{Dt-i})$ w.r.t. $t$ one gets (as $R\rightarrow R_{GM}$),
\begin{align}
E &=1-\eta^2-\frac{2M}{R(t)} \sim e^{-\epsilon t/R_{GM}} \label{D-i3}.
\end{align}
From $eq^n(\ref{D-i3})$ we see that at late times, $1-\eta^2-\frac{2M}{R(t)} \sim e^{-\epsilon t/R_{GM}}$. Since we are interested in the incipient limit, that is, in late times of the collapsing process, we can choose the behaviour of $R(t)$ at early times as per our convenience for simplifying the calculations. Therefore, we choose both past and future behaviour of $R(t)$ to be stationary. Hence, we can take the metric to be flat for all $t\in(-\infty,0)$. Stationarity in future can be achieved by taking a cut-off time $t_f$ for the collapse and then allowing $t_f\rightarrow\infty$, thus going into the continual collapse case till the black hole is formed. Therefore,
\begin{align}
E&= 
\begin{cases}
1, \ \ \ \ \ \ \ \ \ \ \ \ \ \ \ \ \ for \ \ \ t\in(-\infty,0) & \\
e^{-\epsilon t/R_{GM}}, \ \ \ \ for \ \ \ t\in(0,t_f) & \\
e^{-\epsilon t_f/R_{GM}}, \ \ \ for \ \ \ t\in(t_f,\infty). &
\end{cases} \label{D-choice1}
\end{align} 
The above choice of $R(t)$ may seem quite problematic as $\frac{dR}{dt}$ is discontinuous at $0$ and $t_f$, but references \cite{stojkovic1, greenwood1} show that the particle production by the collapsing shell happens in the range, $0<t<t_f$ and in the $t_f\rightarrow\infty$ regime, all the solutions obtained are smooth and well-behaved. Therefore with the above considerations, the wavefunction $\psi$ would capture the whole collapse scenario, and in the limit of $t_f\rightarrow\infty$ or $R(t)\rightarrow R_{GM}$, black hole formation sets in.\\

We note that, at early times, $t\in(-\infty,0)$, the spacetime is Minkowski and therefore the initial vacuum states at $\mathcal{J^{-}}$ (past null infinity) are\footnote{One may see that the intuition behind identifying the states at $\mathcal{J^{-}}$ with the states at $t\in(-\infty,0)$ actually comes from the fact that the observer is at $r\rightarrow\infty$. Now at an early time he is at $t\rightarrow -\infty$ which is $\mathcal{J^{-}}$ and at late times he is at $t\rightarrow\infty$ which is $\mathcal{J^{+}}$.}  just the simple harmonic oscillator ground states (this can be seen from the form of $eq^n(\ref{Schr-sho})$, which with $\widetilde{\eta}=0$, is the SHO equation). Thus,
\begin{align}
\psi_0(b) := \psi(b, \widetilde{\eta}=0) = \left(\frac{\alpha\omega_0}{\pi}\right)^{1/4}e^{-m\omega_0b^2/2}, \label{phi-gs}
\end{align}
where $\psi_0(b)$ represents the SHO ground state and $\lbrace\psi_n(b)\rbrace$ will denote the SHO basis states at early times.\\
\\
$Eq^n(\ref{phi-gs})$ suggests that $\omega_0$ defined in $eq^n(\ref{omega})$ can be identified with the ground state frequency associated with the initial vacuum state.\\
\\
With the help of $eq^n(\ref{phi-gs})$, the exact solution to $eq^n(\ref{Schr-sho})$ is, 
\begin{align}
\psi(b,\widetilde{\eta}) = e^{i\chi(\widetilde{\eta})}\left[\frac{\alpha}{\pi\zeta^2}\right]^{1/4}\exp\left[i\left(\frac{\zeta_{\widetilde{\eta}}}{\zeta}+\frac{i}{\zeta^2}\right)\frac{\alpha b^2}{2}\right], \label{psi-soln}
\end{align}
where $\zeta$ is the solution of the equation,
\begin{align}
\zeta_{\widetilde{\eta}\widetilde{\eta}} + \omega^2(\widetilde{\eta})\zeta = \frac{1}{\zeta^3}, \label{dif1}
\end{align}
with the following initial conditions,
\begin{align}
&\zeta(0)=\frac{1}{\sqrt{\omega_0}}, \label{dif2} \\
&\zeta_{\widetilde{\eta}} (0) = 0 \label{dif3},
\end{align}
and, $\chi(\widetilde{\eta})$ is given by,
\begin{align}
\chi(\widetilde{\eta}) := -\frac{1}{2}\int^{\widetilde{\eta}}_0\frac{d\widetilde{\eta}^{'}}{\zeta^2(\widetilde{\eta}^{'})}. \label{dif4}
\end{align}
Differential equations of the form $eq^n(\ref{Schr-sho})$ have been extensively studied in \cite{dantas1, lewis1, lewis2, pedrosa1, kolopanis1}.\\ \\

From $eq^n s (\ref{omega})$, (\ref{D-i3}) and (\ref{D-choice1}), we have the following (for $t>0$),
\begin{align}
\omega(\widetilde{\eta}(t)) = e^{\epsilon t/2R_{GM}}\omega_0 .
\label{omega1}
\end{align}
Using \ $eq^n(\ref{eta2})$ and $eq^n(\ref{omega1})$, 
\begin{align}
\Omega(t) &= \left(\left.\frac{\partial\widetilde{\eta}}{\partial t}\right|_{t>0}\right)\omega(\widetilde{\eta}) = e^{-\epsilon t/2R_{GM}}\omega_0 ,
\label{omega2}
\end{align}
where $\Omega(t)$ is defined to be the frequency w.r.t. time $t$.\\

We note that at early times ($\mathcal{J^{-}}$), the states are the initial vacuum states of SHO, as described by $\psi_0(b)$. With time, the frequency of the states $\Omega(t)$ evolve, as per $eq^n(\ref{omega2})$, and more and more states get excited. Finally, when the observer measures them at $\mathcal{J^{+}}$ (future null infinity), that is for some $t\in(t_f,\infty)$, we have the following mode expansion (following the evolution n the Schr\"{o}dinger picture\cite{sakurai1}),
\begin{align}
\psi(b,t) = \sum_n c_n(t)\phi_n(b), \label{psigen}
\end{align}
where $c_n(t)$ represent the probability amplitudes. The final SHO states $\lbrace\phi_n(b)\rbrace$ are with the frequency $\Omega_f=\Omega(t_f)$ (a constant), given by,
\begin{align}
\phi_n(b) = \left(\frac{\alpha\Omega_f}{\pi}\right)^{1/4}\frac{e^{-\alpha\Omega_f b^2/2}}{\sqrt{2^n n!}}H_n(\sqrt{\alpha\Omega_f}b), \label{phin123}
\end{align}
where $H_n$ are the Hermite polynomials. Observe that,
\begin{align}
\Omega(t_f) = e^{-\epsilon t_f/2R_{GM}}\omega_0;
\label{omega3}
\end{align}
$c_n$ can be computed from an overlap integral as (see appendix),
\begin{align}
c_n = \begin{cases} 
\frac{(-1)^{n/2}e^{i\chi}}{(\Omega_f\zeta^2)^{1/4}}\sqrt{\frac{2}{P}}\left(1-\frac{2}{P}\right)^{n/2}\frac{(n-1)!!}{\sqrt{n!}}, \ \ \ \ for \ \ \ even \ n & \\   
0, \ \ \ \ \ \ \ \ \ \ \ \ \ \ \ \ \ \ \ \ \ \ \ \ \ \ \ \ \ \ \ \ \ \ \ \ \ \ \ \ \ \ \ \ for \ \ \ odd \ n, &
\end{cases} \label{cn}
\end{align}
where $P:=1-\frac{i}{\Omega_f}\left(\frac{\zeta_{\widetilde{\eta}}}{\zeta}+\frac{i}{\zeta^2}\right)$.

\section{Unitarity}

\subsection{Density Matrix approach}
We shall now calculate the density matrices, $\hat{\rho}_{i}$ and $\hat{\rho}_{f}$, for the initial ($\mathcal{J^{-}}$) and the final ($\mathcal{J^{+}}$) states respectively. $\hat{\rho}_{i}$ and $\hat{\rho}_{f}$ can be written as (see \cite{saini1, saini2}),
\begin{align}
\hat{\rho}_i &= \sum_{m,n}l_m l_n^{*}|\psi_m\rangle\langle\psi_n|, \label{rhoi} \\
\hat{\rho}_f &= \sum_{m,n}c_m c_n^{*}|\phi_m\rangle\langle\phi_n|, \label{rhof}
\end{align} 
where, $l_n$ and $c_n$ are the probability amplitudes appearing in the intial and final states respectively. \\

Since initially the system was in the SHO eigenstates $\lbrace\psi_n\rbrace$ and the wavefunction was normalized, we obtain,
\begin{align}
Tr(\hat{\rho}_i) = 1. \label{trrhoi}
\end{align}
From $eq^n(\ref{cn})$, with $\lambda := \left|1-\frac{2}{P}\right|$, we have,
\begin{align}
Tr(\hat{\rho}_f) &= \sum_{even \ n} |c_n|^2 \nonumber\\
&= \frac{2}{\sqrt{\Omega_f\zeta^2}|P|}\sum_{even \ n}\frac{(n-1)!!}{n!}\lambda^n \nonumber\\
&= \frac{2}{\sqrt{\Omega_f\zeta^2}|P|}\frac{1}{\sqrt{1-\lambda^2}} \nonumber\\
&= \frac{2}{\sqrt{\Omega_f\zeta^2}|P|}\frac{1}{\sqrt{1-\left|1-\frac{2}{P}\right|^2}}. \label{trrhof1}
\end{align}
$P$ has been computed explicitly and used in $eq^n(\ref{trrhof1})$ to obtain (see appendix),
\begin{align}
Tr(\hat{\rho}_f) = 1. \label{trrhof2}
\end{align}
$eq^n(\ref{trrhof2})$ shows that the necessary condition for the unitary evolution of states holds. For the sufficient condition, we compute $Tr(\hat{\rho}_f^2)$. From $eq^n(\ref{rhof})$, 
\begin{align}
\hat{\rho}_f &= \sum_{m,n}c_m c_n^*|\phi_m\rangle\langle\phi_n| \nonumber\\
leading \ to, \ \hat{\rho}_f^2 &= \left(\sum_{m,n}c_m c_n^*|\phi_m\rangle\langle\phi_n|\right)\left(\sum_{i,j}c_i c_j^*|\phi_i\rangle\langle\phi_j|\right) \nonumber\\
&= \sum_{m,n,i,j} c_m c_i c_n^* c_j^*|\phi_m\rangle\langle\phi_n|\phi_i\rangle\langle\phi_j| \nonumber\\
&= \sum_{m,n,j} c_m c_j^*|c_n|^2|\phi_m\rangle\langle\phi_j| \nonumber\\
&= \sum_{m,j} c_m c_j^*|\phi_m\rangle\langle\phi_j|\left(\sum_n|c_n|^2\right) \nonumber\\
&= \sum_{m,j} c_m c_j^*|\phi_m\rangle\langle\phi_j| \nonumber\\
&\left(as, \left(\sum_n|c_n|^2\right)=1 \ by \ eq^n(\ref{trrhof2})\right) \nonumber\\
&=\hat{\rho}_f. \label{trrho2f}
\end{align}
Therefore, by $eq^n(\ref{trrho2f})$ we get,
\begin{align}
Tr(\hat{\rho}_f^2) = Tr(\hat{\rho}_f) = 1. \label{idemp}
\end{align}
Analytically, we have shown that the idempotency of the final density matrix holds indicating a pure quantum state to pure quantum state transition.

\subsection{Conservation of Probability approach}
The probability current 4-vector $J^\mu$ can be defined as,
\begin{align}
&J^0 = |\psi|^2, \label{j1} \\
&\vec{J} = \frac{1}{2\alpha i}[\psi^{*}\vec{\nabla}\psi - \psi\vec{\nabla}\psi^{*}]. \label{j2}
\end{align}
As $b$ is an eigenfunction of $\mathbf{B}$, it is independent of the spatial coordinates $x^i$. Thus, we conclude that $\vec{J}=\vec{0}$. This further suggests,
\begin{align}
\nabla_\mu J^\mu = \frac{\partial |\psi|^2}{\partial t_{obs}}. \label{j3}
\end{align} 
Writing $t_{obs} = t$ (for the observer's time coordinate), we have (from equation (\ref{eta2})),
\begin{align}
&\nabla_\mu J^\mu =  \frac{\partial |\psi|^2}{\partial t} = \frac{\partial |\psi|^2}{\partial\widetilde{\eta}}\frac{\partial\widetilde{\eta}}{\partial t} = E\frac{\partial |\psi|^2}{\partial\widetilde{\eta}} \nonumber \\
&For, \ R\rightarrow R_{GM}, \ \nabla_\mu J^\mu = 0 \ \ \ (as, \ E\rightarrow 0)
\label{j4}
\end{align}
Again analytically, we have shown from ($eq^n(\ref{j4})$), that probability is conserved in the system, in the incipient limit of black hole formation. 

\section{Conclusion}
We showed analytically and comprehensively that the black hole radiation, for a spacetime which is not asymptotically flat, is processed with a unitary evolution. This is confirmed from the density matrix consideration as well as from the conservation of probability consideration. \\

The Schr\"{o}dinger-like wave equations that we used bear resemblance to a minisuperspace version of Wheeler-DeWitt equations\cite{dewitt}. Interestingly, such equations have a present resurgence, in the context of issues concerned with unitarity\cite{sridip1, sridip2, sridip3}. \\

Saini and Stojkovic\cite{saini1} had showed that black hole radiation is processed with a unitary evolution, for a Schwarzchild black hole, from the density matrix consideration. However, they had achieved their conclusion through numerical estimates. We worked with a more general, metric, the global monopole metric, and results for the Schwarzchild case is recovered from this by putting $\eta=0$. \\

The computations on unitarity are all in the incipient limit, the limit of black hole formation. Hence, it does not really take care of the complete black hole evaporation process. However, if unitarity is preserved in this limit, it should be valid at every instant of time. 
\\

In saying this, we further emphasize that, what we have shown in this paper is that black hole radiation is unitary in a not asymptotically flat background spacetime. The present result of unitarity in spacetime that is not asymptotically flat, together with the results obtained in \cite{das1} that the unitarity is preserved for a Reissner-Nordstrom metric which is not globally hyperbolic, settles the issue of conservation of unitarity in spherically symmetric, static (1+3) dimensional spacetimes of the form as given in $eq^n(\ref{f1})$. It also deserves mention that similar results for a Schwarzschild backround obtained in \cite{saini1} numerically, can be arrived at as a special case from both of these more involved examples. So the results are quite consistent, and should have significant implications towards the resolution of the information loss paradox. \\

\section*{Acknowledgements}
AD would like to thank the Department of Science and Technology, Government of India for providing the INSPIRE-SHE scholarship which helped immensely in this research work.

\section*{Appendix}
\subsection*{\bf Alternate motivation for $S_{shell}$}
In this section, we present a different action than $S_{shell}$. We shall call it $S_{new}$. We shall further show that in the incipient limit it will give rise to $\mathcal{H}_{shell}$ and $\Pi_{shell}$. Since we know that the shell behaves like a relativistic particle, we define the new action to be,
\begin{align}
S_{new} = &-\int d\tau \ M = -\int dT \ \frac{M}{T_\tau}, \nonumber\\
= &-4\pi\sigma\int dT \ R^2\left[1-2\pi\sigma R\sqrt{1-R_T^2}\right] \nonumber\\
&+\int dT \ \frac{\eta^2 R}{2}\sqrt{1-R_T^2}, \nonumber\\
= &-4\pi\sigma\int dt \ R^2\left[\sqrt{E-\frac{1-E}{E}R_t^2}-2\pi\sigma R\sqrt{E-\frac{R_t^2}{E}}\right] \nonumber\\ 
&+\int dt \ \frac{\eta^2 R}{2}\sqrt{E-\frac{R_t^2}{E}}\label{s1}.
\end{align} 
Then,
\begin{align}
\mathcal{L}_{new} = &-4\pi\sigma R^2\left[\sqrt{E-\frac{1-E}{E}R_t^2}-2\pi\sigma R\sqrt{E-\frac{R_t^2}{E}}\right] \nonumber\\
&+\frac{\eta^2 R}{2}\sqrt{E-\frac{R_t^2}{E}} \label{lnew}, \\
\Pi_{new} = & \ \frac{\partial\mathcal{L}_{new}}{\partial R_t} \nonumber\\
= & \ \frac{4\pi\sigma R^2R_t}{\sqrt{E}}\left[\frac{1-E}{\sqrt{E^2-(1-E)R_t^2}}-\frac{2\pi\sigma R}{\sqrt{E^2-R_t^2}}\right] \nonumber\\
& - \frac{\eta^2 R}{2}\frac{R_t}{\sqrt{E}\sqrt{E^2-R_t^2}} \label{pnew},\\
\mathcal{H}_{new} = & \ \Pi_{new}R_t - \mathcal{L}_{new} \nonumber\\
= & \ 4\pi\sigma E^{3/2}R^2\left[\frac{1}{\sqrt{E^2-(1-E)R_t^2}} - \frac{2\pi\sigma R}{\sqrt{E^2-R_t^2}}\right] \nonumber\\
&-\frac{\eta^2 R}{2}\frac{E^{3/2}}{\sqrt{E^2-R_t^2}} \label{hnew}. 
\end{align}
In the incipient limit we have,
\begin{align}
&\mathcal{H}_{new} = \frac{4\pi E^{3/2}\mu R^2}{\sqrt{E^2-R_t^2}} \label{hi}, \\
&\Pi_{new} = \frac{4\pi\mu R^2 R_t}{\sqrt{E}\sqrt{E^2-R_t^2}} \label{pi}, 
\end{align}
where, $\mu:=\sigma\left(1-2\pi\sigma R_{GM}-\frac{\eta^2}{8\pi\sigma R_{GM}}\right)$. Observe that these are the exact same equations we had obtained before in this incipient limit.

\subsection*{\bf Computation of $c_n$}
Now let us compute the $c_n$'s explicitly. We know that,
\begin{align}
\psi(b,t) = \sum_n c_n(t)\phi_n(b), \label{psigen}
\end{align}
From the overlap integral we have,
\begin{align}
c_n = &\int db \ \phi_n^{*}\psi = \left(\frac{\alpha^2\Omega_f}{\pi^2\zeta^2}\right)^{1/4}\frac{e^{i\chi(\widetilde{\eta})}}{\sqrt{2^n n!}} \nonumber\\
&\int db \ \exp\left[-\frac{\alpha\Omega_f b^2}{2}+i\left(\frac{\zeta_{\widetilde{\eta}}}{\zeta}+\frac{i}{\zeta^2}\right)\frac{\alpha b^2}{2}\right]H_n\left(\sqrt{\alpha\Omega_f}b\right) \label{cn1}, \\
= &\left(\frac{1}{\Omega_f\pi^2\zeta^2}\right)^{1/4}\frac{e^{i\chi(\widetilde{\eta})}}{\sqrt{2^n n!}} \nonumber\\
&\int dx \ \exp\left[-\frac{x^2}{2}+\frac{x^2}{2}\frac{i}{\Omega_f}\left(\frac{\zeta_{\widetilde{\eta}}}{\zeta}+\frac{i}{\zeta^2}\right)\right]H_n(x) \nonumber\\ 
&(with, \ x:=\sqrt{\alpha\Omega_f}b),  \nonumber\\
= &\left(\frac{1}{\Omega_f\pi^2\zeta^2}\right)^{1/4}\frac{e^{i\chi(\widetilde{\eta})}}{\sqrt{2^n n!}}\int dx \ e^{-Px^2/2} H_n(x) \nonumber\\
&\left(with, \ P:=1-\frac{i}{\Omega_f}\left(\frac{\zeta_{\widetilde{\eta}}}{\zeta}+\frac{i}{\zeta^2}\right)\right) \label{cn2} \\
= &\left(\frac{1}{\Omega_f\pi^2\zeta^2}\right)^{1/4}\frac{e^{i\chi(\widetilde{\eta})}}{\sqrt{2^n n!}} I_n \nonumber\\ 
&\left(with, \ I_n:=\int dx \ e^{-Px^2/2} H_n(x)\right) \label{cn3}.
\end{align}
To compute $I_n$, let us consider the following generating function for the $H_n(x)$,
\begin{align}
&J(z) = \int dx \ e^{-Px^2/2}e^{-z^2+2zx} = \sqrt{\frac{2\pi}{P}}e^{-z^2(1-2/P)}, \nonumber\\
since, \ &e^{-z^2+2zx} = \sum_{n=0}^\infty\frac{z^n}{n!}H_n(x), \nonumber\\
&\int dx \ e^{-Px^2/2}H_n(x) = \left.\frac{d^n}{dz^n}J(z)\right|_{z=0}, \nonumber\\
thus, \ &I_n = \sqrt{\frac{2\pi}{P}}\left(1-\frac{2}{P}\right)^{n/2}H_n(0), \nonumber\\
as, \ &H_n(0) = \begin{cases}
(-1)^{n/2}\sqrt{2^n n!}\frac{(n-1)!!}{\sqrt{n!}}, \ \ \ \ for \ \ \ even \ n & \\
0, \ \ \ \ \ \ \ \ \ \ \ \ \ \ \ \ \ \ \ \ \ \ \ \ \ \ \ \ \ for \ \ \ odd \ n. &
\end{cases}\nonumber
\end{align}
Thus we have,
\begin{align}
&c_n = \begin{cases} 
\frac{(-1)^{n/2}e^{i\chi}}{(\Omega_f\zeta^2)^{1/4}}\sqrt{\frac{2}{P}}\left(1-\frac{2}{P}\right)^{n/2}\frac{(n-1)!!}{\sqrt{n!}}, \ \ \ \ for \ \ \ even \ n & \\   
0, \ \ \ \ \ \ \ \ \ \ \ \ \ \ \ \ \ \ \ \ \ \ \ \ \ \ \ \ \ \ \ \ \ \ \ \ \ \ \ \ \ \ \ \ for \ \ \ odd \ n. &
\end{cases} \label{cn4} 
\end{align}

\subsection*{\bf Explicit computation of $Tr(\hat{\rho}_f)$}
We know that,
\begin{align}
Tr(\hat{\rho}_f) = \frac{2}{\sqrt{\Omega_f\zeta^2}|P|}\frac{1}{\sqrt{1-\left|1-\frac{2}{P}\right|^2}} \label{tr1}.
\end{align}
To calculate $P$ explicitly, let us give the solution of, 
\begin{align}
\zeta_{\widetilde{\eta}\widetilde{\eta}} + \omega^2(\widetilde{\eta})\zeta = \frac{1}{\zeta^3}, \label{dif1}
\end{align}
as,
\begin{align}
&\zeta = \frac{1}{\sqrt{\omega_0}}\sqrt{\widetilde{\epsilon}^2+\varepsilon^2} \label{z1}, \\
&\zeta_{\widetilde{\eta}} = \frac{1}{\omega_0\zeta}(\widetilde{\epsilon}\widetilde{\epsilon}_{\widetilde{\eta}}+\varepsilon\varepsilon_{\widetilde{\eta}}) \label{z2}, 
\end{align}
where in terms of Bessel's functions, we have,
\begin{align}
&\widetilde{\epsilon} = \frac{\pi u_0}{2}[Y_0(2\omega_0)J_1(u_0)-J_0(2\omega_0)Y_1(u_0)] \label{d1}, \\
&\varepsilon = \frac{\pi u_0}{2}[Y_1(2\omega_0)J_1(u_0)-J_1(2\omega_0)Y_1(u_0)] \label{d2}, \\
&\widetilde{\epsilon}_{\widetilde{\eta}} = -\pi\omega_0^2[Y_0(2\omega_0)J_0(u_0)-J_0(2\omega_0)Y_0(u_0)] \label{d3}, \\
&\varepsilon_{\widetilde{\eta}} = -\pi\omega_0^2[Y_1(2\omega_0)J_0(u_0)-J_1(2\omega_0)Y_0(u_0)] \label{d4},
\end{align}
where $u_0:=2\omega_0\sqrt{1-\widetilde{\eta}}$.\\
\\
Now substituting the definition of $P$ ($eq^n(\ref{cn2})$) in $eq^n(\ref{tr1})$, we have (using {\itshape{Mathematica}}),
\begin{align}
Tr(\hat{\rho}_f) = \frac{|\zeta^2\Omega_f|}{\sqrt{\zeta^2\Omega_f}\sqrt{-\Im[\zeta^2\Omega_f]\Re[\zeta\zeta_{\widetilde{\eta}}]+(1+\Im[\zeta\zeta_{\widetilde{\eta}}])\Re[\zeta^2\Omega_f]}} \label{trmath}.
\end{align}
$\left.\right.$\\
Now as $\Omega_f$, $\zeta$ and $\zeta_{\widetilde{\eta}}$ are real (as is evident from $eq^ns(\ref{z1}-\ref{d4})$) , we get from $eq^n(\ref{trmath})$,
\begin{align}
Tr(\hat{\rho}_f) = 1 \label{trf}.
\end{align}

\end{document}